\documentclass{revtex4}

\usepackage{graphicx}
\begin{document}
\bibliographystyle{revtex}

\def\ba{\begin{eqnarray}}
\def\ea{\end{eqnarray}}
\def\bq{\begin{equation}}
\def\eq{\end{equation}}
\def\lsim{\mathrel{\raisebox{-.6ex}{$\stackrel{\textstyle<}{\sim}$}}}
\def\gsim{\mathrel{\raisebox{-.6ex}{$\stackrel{\textstyle>}{\sim}$}}}
\newcommand{\sla}[1]{/\!\!\!#1}
\def\fbi{~fb$^{-1}\;$}

\title{Higgs couplings at the LHC}



\author{Dieter Zeppenfeld}
\email[dieter@pheno.physics.wisc.edu]{} 
\affiliation{Department of Physics, University of Wisconsin, 
Madison, WI 53706, USA}


\date{\today}

\begin{abstract}
The observation of a SM-like Higgs boson in multiple channels at the LHC
allows the extraction of Higgs couplings to gauge bosons and fermions.
The precision achievable at the LHC, for an integrated luminosity of 
200~fb$^{-1}$, is reviewed and updated.
\end{abstract}

\maketitle

\section{Coupling determination at the LHC}

One of the prime tasks of the LHC will be to probe the mechanism of 
electroweak gauge symmetry breaking. 
Beyond observation of the various CP even and CP odd scalars 
which nature may have in store for us~\cite{ATLAS,CMStp,LHCHiggsI},
this means the determination of 
the couplings of the Higgs boson to the known fermions and gauge bosons,
i.e. the measurement of $Htt$, $Hbb$, $H\tau\tau$ and $HWW$, $HZZ$, 
$H\gamma\gamma$, $Hgg$ couplings, to the extent possible. 

Clearly this task very much depends on the expected Higgs boson mass. For
$m_H>200$~GeV and within the SM, only the $H\to ZZ$ and $H\to WW$ channels 
are expected to be observable, and the two gauge boson modes are related 
by SU(2). A much richer spectrum of decay modes
is predicted for the intermediate mass range, i.e. if a SM-like
Higgs boson has a mass between the LEP2 limit of 114~GeV and 
the $Z$-pair threshold. The main reasons for focusing on this range are 
present indications from electroweak precision
data, which favor $m_H\lsim 200$~GeV~\cite{EWfits}, as well as expectations
within the MSSM, which predicts the lightest Higgs boson to have a mass
$m_h\lsim 135$~GeV. In this contribution I 
update and extend recent predictions for
Higgs coupling measurements at the LHC~\cite{zknr} for a Higgs boson
with couplings qualitatively similar to the SM case.

\section{Survey of intermediate mass Higgs channels}

\begin{figure}
 \includegraphics[width=9.0cm, angle=90]{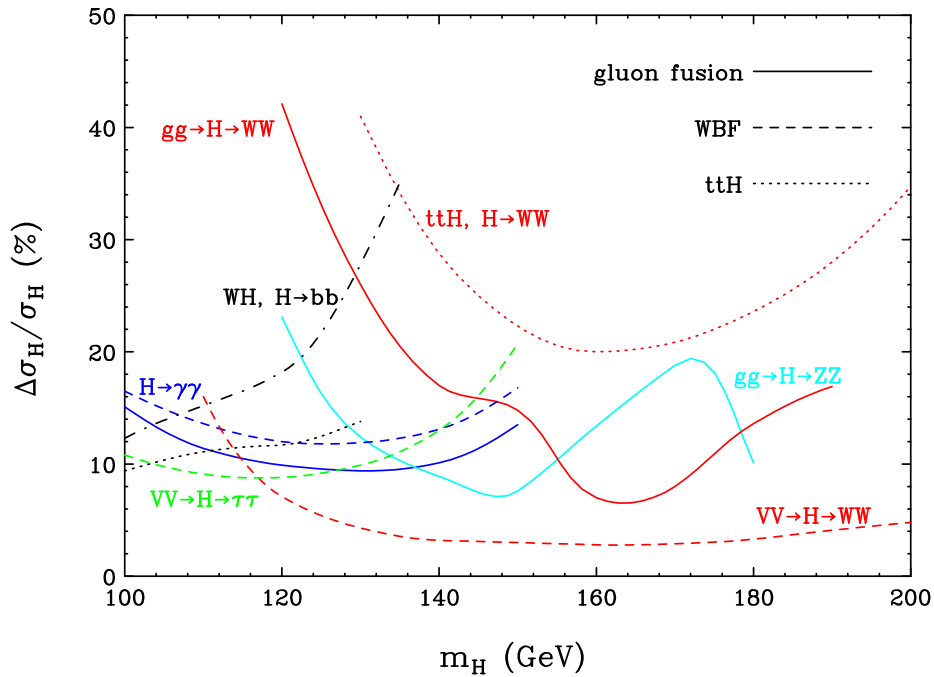}%
 \caption{Expected relative error on the determination of $B\sigma$ for
various Higgs search channels at the LHC with 200~fb$^{-1}$ of data. Solid 
lines are for inclusive Higgs production channels which are dominated by gluon
fusion. Expectations for weak boson fusion are given by the dashed lines.
The black-dotted line is for $ttH,H\to b\bar b$ as analyzed in 
Ref.~\cite{Drollinger:2001ym}. The $ttH,H\to W^+W^-$~\cite{Maltoni:2002jr}
(red dotted) and $WH,H\to b\bar b$~\cite{Drollinger:2002uj} (dash-dotted) 
curves assume 300~fb$^{-1}$ of data and high luminosity running.}
 \label{fig:delsigh}
 \end{figure}

The total production cross section for a SM Higgs boson at the LHC 
is dominated by the gluon fusion process, $gg\to H$, which largely 
proceeds via a top-quark loop. Thus, inclusive Higgs searches will 
collectively be called ``gluon fusion'' channels in the following. 
Three inclusive channels are highly promising for the SM Higgs boson 
search~\cite{ATLAS,CMStp,LHCHiggsI},
\ba 
\label{eq:ggHAA}
gg\to H\to \gamma\gamma\;, \qquad 
{\rm for}\quad m_H \lsim 150~{\rm GeV}\;, \\
\label{eq:ggHZZ}
gg\to H\to ZZ^*\to 4\ell\;,\qquad 
{\rm for}\quad m_H \gsim 120~{\rm GeV}\;, 
\ea
and
\bq
\label{eq:ggHWW}
gg\to H\to WW^*\to \ell\bar\nu\bar\ell\nu\;, \qquad 
{\rm for}\quad m_H \gsim 130~{\rm GeV}\;. 
\eq
The $H\to \gamma\gamma$ signal can be observed as a narrow and high 
statistics $\gamma\gamma$ invariant mass peak, albeit on a very large diphoton 
background. A few tens of $H\to ZZ^*\to 4\ell$ events are expected to be 
visible in 100~fb$^{-1}$ of data, with excellent signal to background 
ratios (S/B), ranging between 1:1 and 6:1, in a narrow four-lepton invariant 
mass peak. Finally, the $H\to WW^*\to \ell\bar\nu\bar\ell\nu$ mode is visible
as a broad enhancement of event rate in a 4-lepton transverse mass 
distribution, with S/B between 1:4 and 1:1 (for favorable values of the 
Higgs mass, around 170~GeV). 
Analyses of these inclusive channels have been performed by CMS and ATLAS
at the hadron level and include full detector simulations. These complete
analyses were used as input in Ref.~\cite{zknr}.  Expected accuracies for
the three inclusive channels are shown as solid lines in 
Fig.~\ref{fig:delsigh}.

Additional and crucial information on the Higgs boson
can be obtained by isolating Higgs production in weak boson fusion (WBF), i.e.
by separately observing $qq\to qqH$ and crossing related processes, in which
the Higgs is radiated off a $t$-channel $W$ or $Z$.
Specifically, it was shown in parton 
level analyses that the weak boson fusion channels, with subsequent Higgs
decay into photon pairs~\cite{RZ_gamgam,R_thesis},
\bq
\label{eq:wbfHAA}
qq\to qqH,\;H\to \gamma\gamma\;, \qquad 
{\rm for}\quad m_H \lsim 150~{\rm GeV}\;,
\eq
into $\tau^+\tau^-$ pairs~\cite{R_thesis,RZ_tautau_lh,RZ_tautau_ll},
\bq
\label{eq:wbfHtautau}
qq\to qqH,\;H\to \tau\tau\;, \qquad 
{\rm for}\quad m_H \lsim 150~{\rm GeV}\;,
\eq
or into $W$ pairs~\cite{R_thesis,RZ_WW,KPRZ}
\bq
\label{eq:wbfHWW}
qq\to qqH,\;H\to WW^*\to e^\pm \mu^\mp /\!\!\!{p}_T\;, 
\qquad {\rm for}\quad m_H \gsim 110~{\rm GeV}\;,
\eq
can be isolated at the LHC. 
The weak boson fusion channels utilize the 
significant background reductions which are expected from 
double forward jet tagging and central jet vetoing techniques, and promise low
background environments in which Higgs decays can be studied in detail.

\begin{table*}[t]
\vspace{0.2in}
\caption{Number of events expected for 
$qq\to qqH,\;H\to WW^*\to ll'\sla p_T$ in 200~fb$^{-1}$ of data, 
and corresponding backgrounds. Predictions for $m_H\ge 150$~GeV include 
$H\to WW^*\to\mu^\pm e^\mp\sla p_T$ decays only~\protect\cite{RZ_WW}.
For smaller Higgs boson masses,
$H\to WW^*\to\mu^+\mu^-\sla p_T,\;e^+e^-\sla p_T$ decays are considered
in addition (see Ref.\protect\cite{KPRZ}).
The expected relative statistical error on the signal cross section, determined
as $\sqrt{N_S+N_B}/N_S$, is given in the last line.}
\vspace{0.15in}
\label{table:WBF.WW}
\begin{tabular}{c|ccccc|ccccc}
$m_H$ &  
110  &  115  &  120  &  130  &  140  &  150  &  160  &  170  &  180  &  190 \\
\hline
$N_S$ &  
102  &  188  &  324  &  740  & 1226  &  908  & 1460  & 1436  & 1172  &  832 \\
$N_B$ &  
164  &  188  &  208  &  254  &  300  &  216  &  240  &  288  &  300  &  324 \\
$\Delta\sigma_H/\sigma_H$ &
16.0\%&10.3\%& 7.1\% & 4.3\% & 3.2\% & 3.7\% & 2.8\% & 2.9\% & 3.3\% & 4.1\% \\
\end{tabular}
\end{table*}

Compared to the analysis of Ref.~\cite{zknr}, new results for 
$H\to WW^*\to l^+\nu {l'}^-\bar\nu$ have become available~\cite{KPRZ} and
will be included in the following. Table~\ref{table:WBF.WW} summarizes
expected event rates (after cuts and including efficiency 
factors) for the combined $qq\to qqH,\;H\to WW^*\to ll'\sla p_T$ channels.
The rates and ensuing statistical errors of the signal cross section 
are given for 100~fb$^{-1}$ of data collected in both the ATLAS and the 
CMS detector. 

The results used in this analysis for the WBF channels were derived 
at the parton level.
First hadron level analyses with full detector simulation 
qualitatively confirm the parton level results~\cite{wbfexp}, 
but yield somewhat lower rates. This can partially be explained 
by initial and final state radiation which produces 
additional jet activity, leading to misidentified forward tagging 
jets. A full 
simulation of forward jets confirms previous assumptions on jet reconstruction
efficiencies which were based on a fast detector simulation. Lower 
efficiencies are found in the very far forward region only, which 
contributes little to the signal.
The parton level results include jet reconstruction efficiencies of 0.86 
per jet. 
Another effect of additional hadronic activity in the full simulation
is a somewhat reduced reconstruction efficiency for isolated leptons,
which was assumed to be 0.95 per lepton in the parton level studies. 
Finally, central jet veto efficiencies, as calculated in PYTHIA versus the
parton level, approximately agree for the signal but show discrepancies for
multi-jet QCD backgrounds, perhaps because PYTHIA uses $2\to 2$ processes 
for the simulation of hard matrix elements.
In view of these unresolved issues, the agreement between parton level and
full simulation results, at the factor 2 level or better, is reassuring.
Also, the parton level analysis 
did not make use of all decay channels (e.g. no $\tau^+\tau^-\to e^+e^-,\;
\mu^+\mu^- +\sla p_T$ decays), the full detector simulation for $H\to WW^*$
has not yet been optimized for $m_H\lsim 140$~GeV, and no analysis has 
yet exploited multivariate techniques 
to enhance the Higgs signals over backgrounds. In light of this, the parton 
level results on WBF processes appear to be quite realistic and I use 
them in the following.  Expected accuracies for
the three WBF channels are shown as dashed lines in 
Fig.~\ref{fig:delsigh}.

Among the associated production channels, $t\bar tH$ production appears 
most promising. Recent analyses have significantly improved the techniques
for observing the decays into 
$b\bar b$~pairs~\cite{Drollinger:2001ym,Green:2001gh},
\bq
\label{eq:ttHbb}
gg,q\bar q\to t\bar tH,\;H\to b\bar b;, \qquad 
{\rm for}\quad m_H \lsim 130~{\rm GeV}\;,
\eq
and $W^+W^-$ pairs~\cite{Maltoni:2002jr},
\bq
\label{eq:ttHWW}
gg,q\bar q\to t\bar tH,\;H\to W^+W^-\;, \qquad 
{\rm for}\quad m_H \gsim 140~{\rm GeV}\;.
\eq
The $H\to b\bar b$ signal in $t\bar tH$ events will be visible with good
purity, S/B~$\approx 2/3$~\cite{Drollinger:2001ym}, 
and the Higgs invariant mass peak allows for a 
direct measurement of backgrounds in the sidebands. The $H\to W^+W^-$ signal
is expected to yield similar purity, but is more difficult to measure
precisely, because the Higgs mass peak cannot be reconstructed. This makes 
precise background determinations challenging.
A third associated production channel which has been re-analyzed recently
is $WH$ production~\cite{Drollinger:2002uj},
\bq
\label{eq:WHbb}
q\bar q\to WH,\;H\to b\bar b\;, \qquad 
{\rm for}\quad m_H \lsim 120~{\rm GeV}\;.
\eq
The particular importance of this channel is that it allows to isolate 
the $H\to b\bar b$ partial width, because the Higgs coupling to $W$'s can be 
separately determined in WBF.

The statistical accuracy with which the signal cross sections of the 
processes in Eqs.~(\ref{eq:ggHAA}-\ref{eq:WHbb}) can be determined is
shown in Fig.~\ref{fig:delsigh}. For 100~fb$^{-1}$ of data per experiment
we expect typical statistical errors of order 10\%. Experimental systematic 
errors, e.g. luminosity errors or knowledge of detector acceptance will 
be substantially smaller, of order 5\% or less. This means that higher
luminosity running can improve experimental errors substantially, provided
that problems associated with pile-up can be overcome. Such dedicated analyses 
have not been finalized yet for all processes. In a conservative approach,
the results below are based on a nominal integrated luminosity of 
200~fb$^{-1}$, unless stated otherwise.

\begin{figure}[thb]
\begin{center}
\includegraphics[width=9.0cm,angle=90]{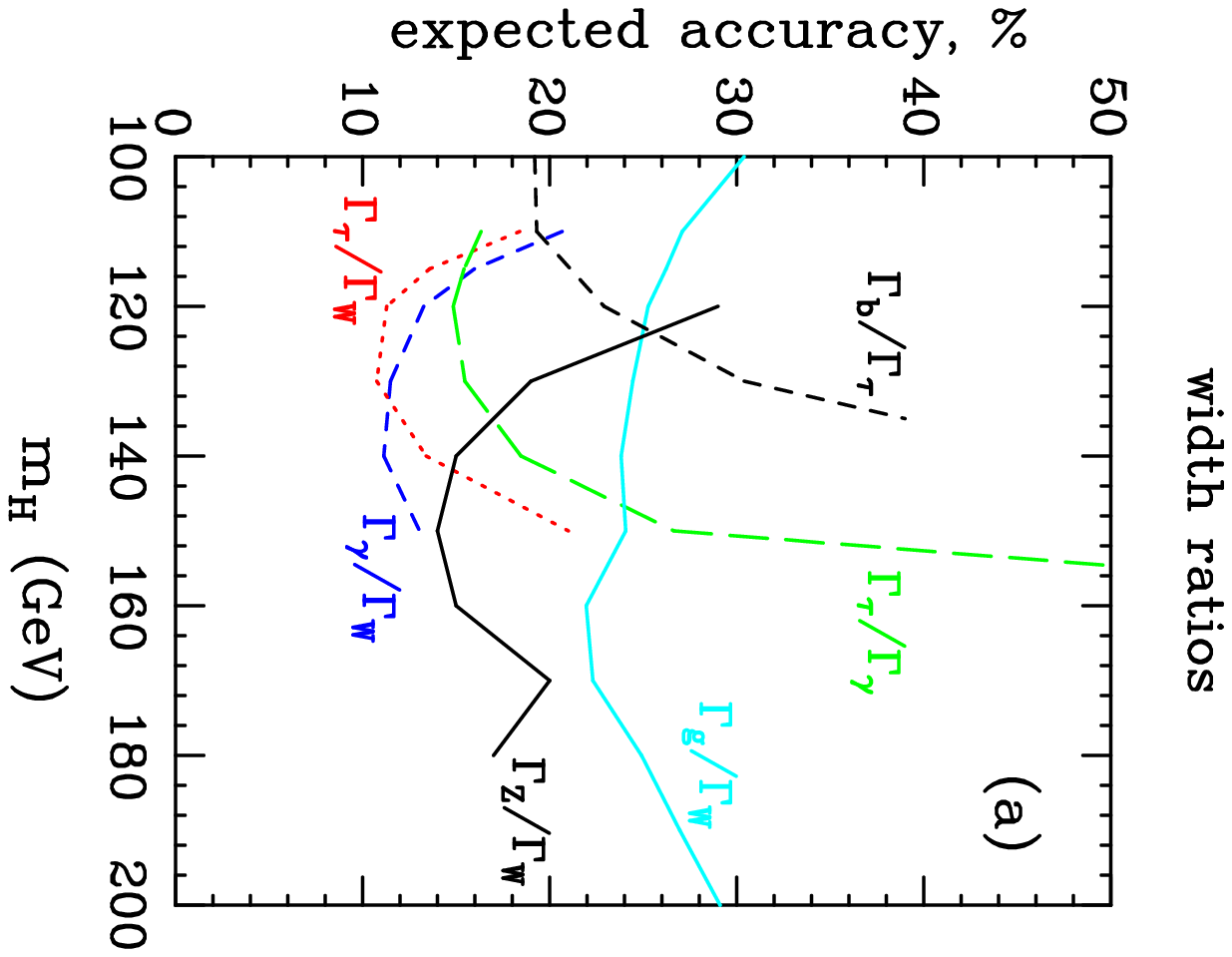}
\includegraphics[width=9.0cm,angle=90]{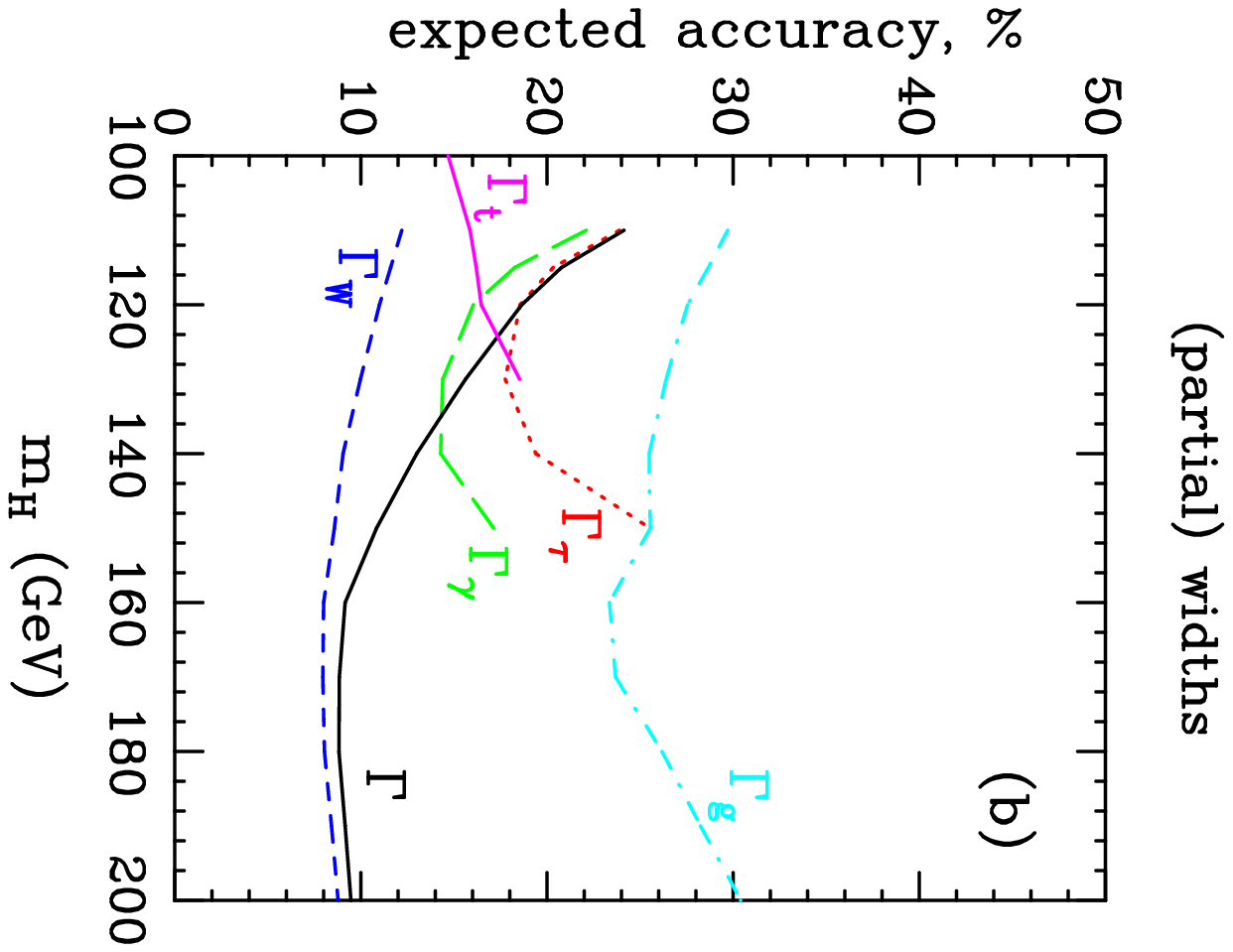}
\end{center}
\caption{Relative accuracy expected at the LHC with 200~fb$^{-1}$ of data for
(a) various ratios of Higgs boson partial widths and (b) the indirect 
determination of partial and total widths $\tilde\Gamma$ and 
$\tilde\Gamma_i=\Gamma_i(1-\epsilon)$.
Width ratio extractions only assume $W,Z$ universality, which can be tested 
at the 15 to 30\% level (solid line). Indirect width
measurements assume $b,\tau$ universality in addition and require a
small branching ratio $\epsilon$ for unobserved modes like $H\to c\bar c$.
(See text).}
\label{fig:Gamma}
\end{figure}

\section{Measurement of Higgs properties}
\label{sec3}

In order to translate the cross section measurements of the various
Higgs production and decay channels into measurements of Higgs boson 
properties, in particular into measurements of the various Higgs boson
couplings to gauge fields and fermions, it is convenient to rewrite 
them in terms of partial widths of various Higgs boson decay channels. 
The Higgs-fermion couplings $g_{Hff}$, for example, which in the SM are 
given by the fermion masses, $g_{Hff} = m_f(m_H)/v$, can be traded for 
$\Gamma_f = \Gamma(H\to \bar ff)$, where, for top-quarks, the final fermions
$f$ would be virtual. Similarly, the 
square of the $HWW$ coupling ($g_{HWW}=gm_W$ in the SM) or the $HZZ$ 
coupling is proportional to the partial widths $\Gamma_W=\Gamma(H\to WW^*)$ 
or $\Gamma_Z=\Gamma(H\to ZZ^*)$.
$\Gamma_\gamma=\Gamma(H\to\gamma\gamma)$ and $\Gamma_g=\Gamma(H\to gg)$
determine the squares of the effective $H\gamma\gamma$ and $Hgg$ couplings.
The Higgs production cross sections are governed by the same squares of
couplings, hence, $\sigma(VV\to H) \sim \Gamma_V$ (for $V=g,\;W,\;Z$).
Combined with the branching fractions $B(H\to ii)=\Gamma_i/\Gamma$
the various signal cross sections measure different combinations of Higgs
boson partial and total widths, $\Gamma_i\Gamma_j/\Gamma$. 

The production rate for WBF is a mixture of $ZZ\to H$ and $WW\to H$ 
processes, and we cannot distinguish between the two experimentally, at the 
LHC. In a large class of models the ratio of $HWW$ and $HZZ$ 
couplings is identical to the one in the SM, however, and this includes the
MSSM. Let us therefore assume that 1) the $H\to ZZ^*$ and $H\to WW^*$ partial 
widths are related by SU(2) as in the SM, i.e. their ratio, $z$, is given 
by the SM value, $z=\Gamma_Z/\Gamma_W=z_{SM}$.
This assumption can be tested, at the 15-20\% level for 
$m_H>130$~GeV, e.g. by forming the ratio 
$B\sigma(gg\to H\to ZZ^*)/B\sigma(gg\to H\to WW^*)$~\cite{zknr}. The expected
precision, as a function of $m_H$, is given by the solid black line in 
Fig.~\ref{fig:Gamma}.

With  $W,Z$-universality, the three weak boson fusion cross sections give us
direct measurements of three combinations of (partial) widths,
\ba
X_\gamma = {\Gamma_W\Gamma_\gamma\over \Gamma}\qquad &{\rm from}&\;\;
qq\to qqH,\; H\to\gamma\gamma \;, \\
X_\tau = {\Gamma_W\Gamma_\tau\over \Gamma}\qquad &{\rm from}&\;\;
qq\to qqH,\; H\to\tau\tau \;, \\
X_W = {\Gamma_W^2\over \Gamma}\quad\qquad &{\rm from}&\;\;
qq\to qqH,\; H\to WW^* \;,
\ea
In addition the three gluon fusion channels provide measurements of 
\ba
Y_\gamma = {\Gamma_g\Gamma_\gamma\over \Gamma}\qquad &{\rm from}&\;\;
gg\to H\to\gamma\gamma \;, \\
Y_Z = {\Gamma_g\Gamma_Z\over \Gamma}\qquad &{\rm from}&\;\;
gg\to H\to ZZ^* \;, \\
Y_W = {\Gamma_g\Gamma_W\over \Gamma}\qquad &{\rm from}&\;\;
gg\to H\to WW^* \;.
\ea
Finally, the $t\bar tH$ and $WH$ associated production channels measure
the combinations
\ba
T_b = {\Gamma_t\Gamma_b\over \Gamma}\qquad &{\rm from}&\;\;
gg\to t\bar tH,\; H\to b\bar b\;, \\
T_W = {\Gamma_t\Gamma_W\over \Gamma}\qquad &{\rm from}&\;\;
gg\to t\bar tH,\; H\to WW^* \;, \\
U_b = {\Gamma_W\Gamma_b\over \Gamma}\qquad &{\rm from}&\;\;
q\bar q\to WH,\; H\to b\bar b \;.
\ea

When extracting Higgs couplings, the QCD uncertainties of production cross
sections enter. These can be estimated via the residual scale dependence
of the NLO predictions and are small (below 5\%) for the WBF 
case~\cite{wbfNLO}, while larger uncertainties are found for gluon 
fusion~\cite{Djouadi:1991tk,Spira:1995rr}. 
Recently, the NNLO QCD corrections for the gluon fusion cross section
have been determined~\cite{HggNNLO}, in the heavy top-quark limit, 
which provides an excellent approximation for the intermediate mass 
Higgs boson considered here. First results indicate a stabilization
of the perturbative expansion at two loops. The diminished scale dependence
at NNLO suggests a remaining error due to higher order effects of as little
as $15\%$. The present analysis assumes a theoretical uncertainty of 20\%
in the gluon fusion cross section. For the $t\bar tH$ cross section, which
has been calculated at NLO as well~\cite{ttHNLO,tthNLOq}, the scale
dependence is reduced dramatically at NLO, 
to a level of about $6\%$~\cite{ttHNLO}. For this analysis I conservatively
take a theory error of $10\%$ into account.

A first test of the Higgs sector is provided by taking ratios 
of the $X_i$'s and ratios of the $Y_i$'s. QCD uncertainties, and all 
other uncertainties related to the initial state, like luminosity and pdf 
errors, largely cancel in these ratios.  For example, 
$X_\tau/X_W=\Gamma_\tau/\Gamma_W$ compares the $\tau\tau H$ Yukawa 
coupling with the $HWW$ coupling, while $X_\gamma/X_W$ and $Y_\gamma/Y_W$
determine the analogous ratio, $\Gamma_\gamma/\Gamma_W$ for the loop-induced
photon--Higgs coupling. Expected errors on theses ratios, for 
200~fb$^{-1}$ of data, are shown in Fig.~\ref{fig:Gamma}(a). For Higgs masses 
between 120 and 140 GeV they are in the 10--15\% range.
Accepting an additional systematic error of about 20\%, a measurement 
of the ratio $\Gamma_g/\Gamma_W$, which determines the $Htt$ to $HWW$ 
coupling ratio, can be performed, by measuring the cross section 
ratios $B\sigma(gg\to H\to\gamma\gamma)/\sigma(qq\to 
qqH)B(H\to\gamma\gamma)$
and  $B\sigma(gg\to H\to WW^*)/\sigma(qq\to qqH)B(H\to WW^*)$. 
Comparing $WH$ to WBF cross sections, the ratio 
$U_b/X_\tau=\Gamma_b/\Gamma_\tau$ determines the ratio of b-quark
to tau Yukawa couplings. Because QCD radiative corrections to both WBF 
and Drell-Yan processes are well known, this cross section ratio has a small
QCD error, which is taken as 10\%. Taking the error estimates of 
Ref.~\cite{Drollinger:2002uj}, for 300\fbi in one detector, and $m_H=120$~GeV,
the $\Gamma_b/\Gamma_\tau$ ratio can be determined with an overall uncertainty
of $\pm 23\%$. Since this measurement becomes worse quickly with increasing
$m_H$, it will be replaced below by the assumption of $b,\tau$~universality.

Beyond the measurement of coupling ratios, minimal additional assumptions 
allow an indirect measurement of the total Higgs width. First of all, the 
$\tau$ rate in WBF is measurable with an accuracy of 
order 10\%. The $\tau$ is a third generation fermion with isospin 
$-{1\over 2}$, just like the $b$-quark. In many models, the ratio of their
coupling to the Higgs is given by the $\tau$ to $b$ mass ratio. 
In addition to $W,Z$-universality we thus assume that
(2) $y = \Gamma_b/ \Gamma_\tau = y_{SM}$ and, finally,
(3) the branching ratio for unexpected channels is small, i.e.
$\epsilon = 1-\left[ B(H\to b\bar b)+B(H\to \tau\tau)+B(H\to WW^*)+
B(H\to ZZ^*)+B(H\to gg)+B(H\to \gamma\gamma)\right] \ll 1$. An error
of 7\% is assigned to the $b,\tau$~universality assumption, corresponding
to the uncertainty induced in $y_{SM}$ by the error on the $b$-quark
mass.

With these three assumptions consider the observable
\begin{eqnarray}
\tilde\Gamma_W &=& X_\tau(1+y) + X_W(1+z) + X_\gamma + Y_W 
\nonumber \\
&=& \biggl(\Gamma_\tau+\Gamma_b +\Gamma_W +\Gamma_Z+
\Gamma_\gamma+\Gamma_g\biggr){\Gamma_W\over\Gamma}
=(1-\epsilon)\Gamma_W  \;.
\end{eqnarray}
$\tilde\Gamma_W$ provides a lower 
bound on $\Gamma(H\to  WW^*)=\Gamma_W$. Provided $\epsilon$ is small 
(within the SM and for $m_H\ge 115$~GeV, $\epsilon < 0.03$ and it is dominated 
by $B(H\to c\bar c)$), the determination 
of $\tilde\Gamma_W$ provides a direct measurement of the $H\to WW^*$ 
partial width. Once $\Gamma_W$ has been determined, the total width of the 
Higgs boson is given by
\bq
\label{eq:Gamma_tot}
\Gamma = {\Gamma_W^2\over X_W}={1\over X_W}
\biggl(X_\tau(1+y) + X_W(1+z) + X_\gamma + \tilde X_g \biggr)^2
{1\over (1-\epsilon)^2}\; .
\eq
Similarly, other partial widths can be extracted by combining the 
ratio measurements discussed above with the $\Gamma_W$ determination, e.g.
\bq
\tilde\Gamma_\tau = {\Gamma_\tau\over\Gamma_W}\tilde\Gamma_W = 
{X_\tau\over X_W}\tilde\Gamma_W\;.
\eq
The top quark Yukawa coupling can be determined by the combination
\bq
\tilde\Gamma_t = {T_b\over X_\tau}\tilde\Gamma_W {1\over y_{SM}} =
{\Gamma_t\Gamma_b\over\Gamma_W\Gamma_\tau}\tilde\Gamma_W 
{\Gamma_\tau^{SM}\over \Gamma_b^{SM}}\;.
\eq

Extractions of $\Gamma_\tau$ and the total width require a measurement 
of the $qq\to qqH, H\to WW^*$ cross section,
which is expected to be available for $m_H\gsim 110$~GeV~\cite{KPRZ}, but
suffers from poor statistics close to the limit set by LEP2.
Consequently, errors are sizable for Higgs masses around 110~GeV,
but stay within the 10--20\% range for the most interesting 
Higgs mass region, between 120 and 140~GeV.
Results for these indirect extractions of (partial) widths are shown in 
Fig.~\ref{fig:Gamma}(b) and look highly promising. 

One should note that only a few of these measurements, most notably the 
$H\to gg$ partial width (over the entire Higgs mass range) and the 
extraction of
$\tilde\Gamma_W$ and $\tilde\Gamma$ for $m_H\gsim 150$~GeV, are limited by 
systematic uncertainties, in particular higher order QCD effects. For the
WBF cross sections the assessment of these errors (5\%) is probably too
conservative. This means that substantial improvements are possible,
in principle, with higher luminosity data, in the 115~GeV~$<m_H<150$~GeV 
mass range. Whether pile-up effects do permit such improvements requires 
detailed studies.


\section{Summary}
\label{sec4}

With an integrated luminosity of 100 fb$^{-1}$ per experiment, the LHC can 
measure various ratios of Higgs partial widths, with accuracies of order 
10 to 25\%. This translates into 5 to 10\%
measurements of various ratios of coupling constants. The 
ratio $\Gamma_\tau/\Gamma_W$ measures the coupling of down-type fermions
relative to the Higgs couplings to gauge bosons. To the extent that the
$H\gamma\gamma$ triangle diagrams are dominated by the $W$ loop,
the width ratio $\Gamma_\tau/\Gamma_\gamma$ probes the same relationship.
The fermion triangles leading to an effective $Hgg$ coupling are expected 
to be dominated by the top-quark, thus, $\Gamma_g/\Gamma_W$ probes the 
coupling of up-type fermions relative to the $HWW$ coupling. This top-quark
Yukawa coupling can be probed directly, at the 15 to $20\%$ level for 
$m_H\lsim 130$~GeV, via $t\bar tH$ production with $H\to b\bar b$.
Finally, for any Higgs boson mass in the range left by LEP2,
the absolute normalization of the $HWW$ coupling is accessible
via the extraction of the $H\to WW^*$ partial width
in weak boson fusion.

These measurements test the crucial aspects of the Higgs sector.
The $HWW$ coupling, being linear in the Higgs field, 
identifies the observed Higgs boson as the scalar 
responsible for the spontaneous breaking of $SU(2)\times U(1)$: a 
scalar without a vacuum expectation value does not exhibit such a 
trilinear coupling at tree level. The particular tensor structure of this
SM $HWW$ coupling can be identified as well in WBF~\cite{PRZ}.
The measurement of the ratios $g_{Htt}/g_{HWW}$ and 
$g_{H\tau\tau}/g_{HWW}$ then probes the mass generation of both up and down
type fermions. 

These measurements may be complemented by observation of 
$WH/ZH,\;H\to b\bar b$ production at the Tevatron or the LHC
to remove assumptions on the $g_{Hbb}/g_{H\tau\tau}$ coupling ratios. 
In all, hadron collider data in 
the LHC era are expected to determine the dominant couplings of a SM 
like Higgs boson at the 5--10\% level.


%
%

%
%

\vspace*{-0.1in}
\begin{acknowledgments}
This work was supported in part by WARF and in part by DOE under Grant
No.~DE-FG02-95ER40896.  
\end{acknowledgments}
\vspace*{-0.1in}

\bibliography{P1_zeppenfeld_0716}

\end{document}